\documentclass[a4paper]{article}
\usepackage{interspeech2012,amssymb,amsmath,graphicx}
\title{Plagiarism Detection in Polyphonic Music using Monaural Signal Separation}

\makeatletter
\def\name#1{\gdef\@name{#1\\}}
\makeatother
\name{{\em Soham De$^1$, Indradyumna Roy$^1$, Tarunima Prabhakar$^2$, Kriti Suneja$^3$,}\\
      {\em Sourish Chaudhuri$^4$, Rita Singh$^4$,
      Bhiksha Raj$^4$}}

\address{$^1$Computer Science \& Engineering, Jadavpur University, Kolkata, WB, India. \\
		$^2$DA-IICT, Gandhinagar, GJ, India. \\
		$^3$Electronics \& Communication Engineering, LNM-IIT, Jaipur, RJ, India. \\
               $^4$Language Technologies Institute, Carnegie Mellon University, Pittsburgh, PA, USA. \\
{\small \tt sohamde@ieee.org, indrar.cse.jdvu@gmail.com, tarunima\_prabhakar@daiict.ac.in, } \\ 
	  {\small \tt suneja212@gmail.com, \{schaudhu, rsingh, bhiksha\}@cs.cmu.edu}}

\begin{document}
\maketitle

\begin{abstract}
Given the large number of new musical tracks released each year, automated approaches to plagiarism detection are essential to help us track potential violations of copyright. Most current approaches to plagiarism detection are based on musical similarity measures, which typically ignore the issue of polyphony in music. We present a novel feature space for audio derived from compositional modelling techniques, commonly used  in signal separation, that provides a mechanism to account for polyphony without incurring an inordinate amount of computational overhead. We employ this feature representation in conjunction with traditional audio feature representations in a classification framework which uses an ensemble of distance features to characterize pairs of songs as being plagiarized or not. Our experiments on a database of about 3000 musical track pairs show that the new feature space characterization produces significant improvements over standard baselines.

\end{abstract}

\section{Introduction}
Music-plagiarism is the use or close imitation of another author's music without proper acknowledgement. Every year, vast numbers of new music tracks are released globally, and questionable similarities exist in some sections of music tracks. Aided by the internet, plagiarism is now noticeable {\em globally}, not just across authors, but also across languages and countries. In 2008 alone, 1.4 billion music tracks were sold internationally. This number has since increased to over 1.8 billion. In 2004, the SACEM, an organization that seeks to protect the rights of the original authors, composers and publishers, was able to manually check only a small percentage of registered pieces for potential copyright violations. With such vast numbers of tracks to monitor, the need for automatic techniques for identification of potential copyright violations and detection of music-plagiarism is clear and paramount.

Current approaches to plagiarism detection use techniques based on musical similarity analysis, which emphasizes finding musical pieces in large databases for retrieval. Various feature sets have been proposed for characterization of musical tracks, including the use of pitch contours, loudness, and cepstral features \cite{Ghias1995, Logan2001}. Approaches to computing similarity given the characterizations of two musical tracks use various approaches including $n$-gram based similarity, geometric distances, and string-matching algorithms \cite{Doraisamy2003, Hanna2007}. There are two main issues with adopting similar techniques for plagiarism detection-- first, these approaches typically ignore the issue of polyphony in the recordings, simply using a monophonic approximation instead \cite{Hanna2007}. Polyphonic music can have multiple overlapping notes and is far more complex from an analysis perspective than monophony. Feature extraction techniques traditionally used for monophonic music cannot be expected to do a good job of representing polyphonic characteristics. As a result, these methods have limited success when applied on polyphonic music. Second, unlike similarity computation for retrieval where a system returns a ranked list, a plagiarism detector needs to decide whether a pair of songs are sufficiently similar that one may have been plagiarized from the other. 

In this paper, we present an approach that can effectively deal with these issues. To tackle the problem of polyphony, we present a novel feature set derived from signal separation based on compositional models \cite{Raj2010}. This feature set represents the magnitude spectrum of each frame in a musical data segment as an additive, weighted combination of a set of bases. The bases  are not constrained to be physically meaningful in this work, but such constraints may be applied as well; e.g. each base could represent the different notes that are expected to be present. In this framework, the weights assigned to bases to compose each spectral vector can be used as a feature representation for the audio frame. 

The second problem of requiring a decision, as opposed to a ranked list in case of retrieval, is tackled by formulating the problem as a discriminative classification task. Given a pair of musical segments and various feature characterizations, we compute an ensemble of distance-based features. These computed distances serve as a representation of the {\it pair} of musical pieces. Each pair is a datapoint with a corresponding label that indicates whether one of the songs is plagiarized from the other or not. We use the labels in conjunction with the distance-based features to train a classification model. Now, given a pair of test musical segments, the system can use this model to decide whether one of the segments may have been plagiarized from the other.

While applications aside from plagiarism detection are beyond the scope of this paper, the techniques described can easily be applied to other related tasks. For instance, for the task of retrieving similar segments, given a musical segment ($M_1$), one can query the classification model using $M_1$ and all other musical segments in the database. Such a system could use the distance from the decision boundary as a score to create a ranked similarity list. The use of signal separation-based embedding for musical segments can also be adopted for other music tasks in polyphonic settings.

The remainder of this paper is organized as follows: in Section \ref{sec:problem}, we present a detailed formulation of the problem of music-plagiarism detection. Section \ref{sec:nmf} discusses the feature representations we use for music tracks as well as to characterize pairs of musical segments as plagiarized or not. In Section \ref{sec:experiments}, we describe the dataset used in our experiments and present our results and our discussion of the same. We conclude the paper in Section \ref{sec:conc}.

\section{Problem Formulation}
\label{sec:problem}

In this paper, we approach the task of detection of music plagiarism in a pairwise discriminative classification framework, where given a pair of musical segments, we wish to predict whether the pair are sufficiently similar so as to be considered plagiarized. Each musical track ($x$) is first represented using a set of feature vectors denoted by $F(x) = \{f_1, f_2, ..., f_n\}$. Each of the individual $f_k$ here represent a {\it class} of features, {\it e.g.} Mel-Frequency Cepstral Coefficient features, pitch contour features. 

We then define a set of $n$ distance functions (one for each feature class extracted for a song), $\phi_1, \phi_2,.. \phi_n$, over pairs of such tracks, $x_i$ and $x_j$, where each of these functions $\phi_k$ computes the distance between the feature vectors of the two music tracks for the $k$-th feature class:
\begin{equation}
\phi_k(x_i,x_j) = D(f_k(x_i),f_k(x_j)), \forall k \in [1,2...n]
\end{equation}

where $D$ represents a distance function that is computed over the $k$-th feature classes for each of the music tracks. As we describe in Section \ref{subsec:dtw}, we use an edit-distance measure to compute distance for our task; we note, however, that any other distance metric could be used in this framework, if applied to a different task. The set of distance scores, thus computed, behaves, in effect, as a feature characterization of the degree of difference between the two songs.
\begin{equation}
\mathcal{F}(x_i, x_j) = \{ \phi_1(x_i,x_j), \phi_2(x_i,x_j),.. \phi_n(x_i,x_j)\}
\end{equation}

 At training time, given information about whether the pair of songs, $x_i$, $x_j$, represent a positive instance of plagiarism, we can use the distance based feature set in a supervised classification framework to train a model that can predict plagiarism, given {\it a pair} of music tracks. Let $\bar{w}$ represent the set of weights for the features learnt at training time, and $\mathcal{G}$ represent the function that computes a score for the datapoint ({\it e.g.} $\mathcal{G}(\bar{w}, \mathcal{F}(x_i, x_j)) = \sum_{k=1}^{n} w_k  \mathcal{F}_k $, for linear regression) . We can then obtain a label $L$ for the pair of music tracks using a thresholded classifier score as follows:
\begin{equation}
H(x_i,x_j) = \mathcal{G}(\bar{w}, \mathcal{F}(x_i, x_j)) - \rho
\end{equation}
\begin{equation}
L = \begin{cases}
  +1, & \text{if } H(x_i,x_j)>0, \\
  -1, & \text{if } H(x_i,x_j)\leq0.
\end{cases}
\end{equation}
where $L=+1$ denotes plagiarized. Unlike tasks requiring similarity computation between tracks for search-like applications, our task of detecting plagiarism is different in that we cannot simply pick the few most similar songs as potentially plagiarized, since this would result in a large number of songs that need manual examination. The parameter $\rho$ is therefore used as a threshold in this formulation, and we try to find an optimal value for this parameter so that we make the fewest mistakes.

\section{Feature Representations from compositional models}
\label{sec:nmf}

Our feature-space design is primarily motivated by the fact that most current approaches do not explicitly address the issue of polyphony in recordings, simply using a monophonic approximation instead, while others consider polyphony as a more general multidimensional mathematical issue. While an ideal solution to this problem would involve separating out the multiple notes or voices in the recording into different tracks, this would require information for each track that is not likely to be available.

We introduce a novel feature set that is based on compositional representation of the magnitude spectra. Specifically, we use a non-negative matrix factorization (NMF)-based embedding for the music tracks \cite{Lee1999}, that we expect will account for polyphony much better than the feature sets traditionally used for music representation. These NMF-based features are used in conjunction with traditional feature representations.
In the following, we first describe the NMF-based feature extraction technique for audio, and then briefly describe traditionally used feature sets in Section \ref{subsec:tradfeat}. In Section \ref{subsec:dtw}, we discuss the distance function used to create a characterization for pairs of tracks.

\subsection{NMF features}
\label{subsec:nmffeat}
NMF is a subspace analysis technique which obtains a parts-based representation of data by imposing non-negative constraints \cite{Lee1999}. Given training data, NMF can learn a set of basis vectors so that we can represent any datapoint as a linear weighted non-negative combination of these vectors. We use the magnitude spectra of the audio signals as our data, since they are guaranteed to be non-negative. We can represent $M_t$, a magnitude spectral vector at time $t$ as:
\begin{equation}
M_t = \sum_{i=1}^N \textbf{b}_i w_{i,t}
\end{equation}
where $\textbf{b}_i$ is the $i$-th basis vector and $w_{i,t}$ is the weight of the basis in frame $t$. $N$ is the number of basis vectors.

If we represent the set of basis vectors using matrix $\textbf{B}_N = [ \textbf{b}_1, ..., \textbf{b}_N ]$, the model and the weights using the matrix $[\textbf{W}_N]_{i,t} = w_{i,t}$, we can write the model as,
\begin{equation}
\textbf{M} = \textbf{B} \textbf{W}
\end{equation}

NMF has been applied to various audio tasks, including blind source separation and separation of speech from music \cite{Raj2010,Virtanen2007}. The intuition behind using the NMF formulation for music is that a polyphonic music segment will be composed additively from various notes, and NMF can estimate the contribution of the various notes. Thus, each audio frame can be represented using the NMF technique in the $N$-dimensional basis space, where the weights for the frame correspond to the co-ordinates for the frame in this space. NMF has a significant advantage over dimensionality reduction techniques such as PCA and ICA in that the number of bases used need not be less than the original space, resulting in an overcomplete basis space. For this task, the basis vectors may be thought of as individual notes present in the music.

We use an exemplar-based basis set for our experiments in this paper, where bases are drawn randomly from a collection of spectral vectors for the source (magnitude spectra vectors in the music data, in our case). Such bases, although lacking an intuitive interpretation, have useful theoretical properties \cite{Smaragdis2009}. For alternate tasks, where more information about the notes and instruments is available, one could constrain the basis set to consist of true notes or learn them from audio libraries.

Once the set of bases $\textbf{B}$ is selected, each magnitude spectral vector in the dataset can be represented as a non-negative-weighted combination of the bases. The weights are obtained using an iterative update rule minimizing a generalized Kullback Leibler divergence \cite{Lee2001a} between $\textbf{M}$ and $\textbf{BW}$ as follows:
\begin{equation}
\label{eq7}
\textbf{W} = \textbf{W} \otimes \frac{\textbf{B}^T . [\frac{\textbf{M}}{\textbf{BW}}]}{\textbf{B}^T . \textbf{1}}
\end{equation}
where \textbf{1} is a matrix of ones and the operation $\otimes$ denotes element-wise multiplications. All divisions are element-wise, as well. Weights $\textbf{W}$ are initialized to unity, and we iterate equation \eqref{eq7} to convergence. For each music piece, we now have a sequence of weight vectors $\textbf{W}$ which can be used as a feature representation for the audio in the basis space. These sequences correspond to one {\it feature class}, as described in Section \ref{sec:problem}. We used a 1024-dimensional representation of each audio frame in the data and 64 basis vectors for all experiments reported in this paper.

\subsection{Traditional Representations for Audio}
\label{subsec:tradfeat}
In addition to the features extracted from the NMF, we extract features using traditional means of analysis of the audio content, that describe the temporal and spectral sound structures effectively. We use the F-score for identifying the more discriminative features for the dataset used \cite{Chen2006}. We expect that augmenting these with the NMF-based features will lead to increased efficiency in the detection of similar music documents in polyphonic settings.

In general, perception of structural boundaries in music is mostly influenced by variations in timbre, tonality and rhythm. Timbral features, such as spectral rolloff, which estimates the amount of high frequency of the signal, and zero crossing rate are extracted. Key strength, a tonality feature which indicates the cross-correlation score for each different tonality candidate, is another feature extracted. Other features extracted from the audio include Mel-Frequency Cepstral Coefficients, the relative Shannon entropy, indicating predominant peaks, kurtosis, indicating trends in the audio signal, standard deviation, skewness, and the amplitude envelope. We also use the novelty curve, obtained from convolution along the main diagonal of the similarity matrix using a Gaussian checkerboard. We use this feature set as the baseline for comparison with the enhanced system that also includes the NMF-based features.

\subsection{Distance Features for Pair Characterization}
\label{subsec:dtw}
In the previous subsections, we described the sets of features extracted for each music track. We  use these features to compute distances which we use to characterize pairs of tracks as plagiarized or not. Given the feature representations for the two tracks, we use the Dynamic Time Warping (DTW) algorithm to compute distances between the corresponding feature sequences. Dynamic time warping is a dynamic programming algorithm for efficiently computing the optimal alignment of non-linearly expanded or contracted sequences.

Based on an inspection of the dataset, we observed that the music in a pair of plagiarized pieces usually differed in rhythm, while retaining other properties that made them sound similar. We incorporate this intuition into our distance computation by using a modified version of the DTW algorithm for rhythm-based features, with local weights to prefer insertions or deletions to substitutions to better account for the variation in rhythm. The modification is as follows:
\begin{equation}
D(i,j) = min \begin{cases}
  D(i-1,j-1)+w_{sub}.d(i,j) \\
  D(i,j-1)+w_{del}.d(i,j) \\
  D(i-1,j)+w_{ins}.d(i,j)
\end{cases}
\end{equation}
We make $w_{sub}$ much larger than $w_{del}$ and $w_{ins}$. This follows from the fact that, even though the DTW would effectively match two sequences having the same musical property with varying rhythm, the distance between the two sequences would still be large owing to the numerous insertions and deletions required. Thus, using a lower weight for insertion and deletion would help in bringing out this similarity.

However, this modification has a potential disadvantage in that the warping path might begin to prefer axis-parallel trajectories due to the significantly lower costs of insertions and deletions. This is overcome by constraining the slope of the warping path over short windows to be within a pre-specified range.

\section{Experimental Results}
\label{sec:experiments}

We used the MIRToolBox in MATLAB \cite{Lar:07} to extract the traditional audio-based features described in Section \ref{subsec:tradfeat}.  This was followed by extraction of the NMF features, and the DTW-based distance computation to set up pairwise characterization of tracks, as described in Section \ref{sec:nmf}. These distance features and true class labels for the training data were then used to train a random forest classifier \cite{Breiman2001} with 150 trees.

\subsection{Dataset}
We created a database of 2966 song pairs, comprising of 966 plagiarized (positive data set) and 2000 non-plagiarized (negative data set) song pairs. The positive instances were obtained from music covers and plagiarism lawsuits \cite{Law,ItwoFS} including the Music Copyright Infringement Resource of the UCLA School of Law. The dataset is comprised of music from a wide range of genres and languages. All recordings were resampled to a uniform 16 kHz sampling rate with a frame length of 40 ms. The training and test data were separated randomly to provide a 9:1 train-test split. 

\subsection{Results}
We compare performance of 3 systems on the test data. We use the traditional feature sets described in Section \ref{subsec:tradfeat} as the {\it baseline}, and compare it with performance using the {\it NMF-features only}, as well as an {\it enhanced feature set} using baseline features along with NMF-features. First, a comparison of overall classification accuracy on the test set is shown in Table \ref{table:acc}. We find that the NMF-only system significantly outperforms the system using the baseline feature set, while the enhanced system significantly outperforms both of them.

\begin{figure}[t]
\centering
\includegraphics[height=48mm, width = 90mm]{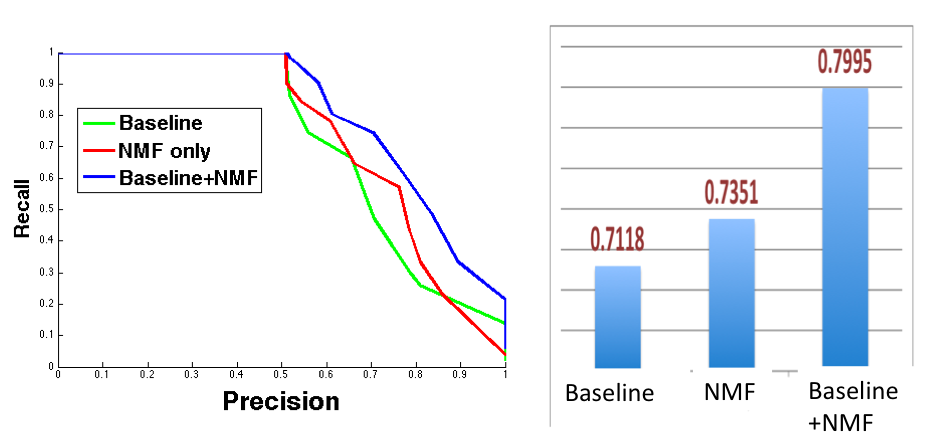}
\caption{\it (L): Precision-Recall ROC curves for the various systems on the plagiarized instances in test data; (R) Corresponding Area Under the Curve (AUC) for the various systems}
\label{fig:roc}
\end{figure}

Figure \ref{fig:roc} compares performance of the systems using ROC plots for precision and recall for the plagiarized instances in the test data. Since precision and recall are both metrics of accuracy, the higher the Area Under the Curve (AUC), the better the performance. We note that the NMF-only outperforms the baseline, but the enhanced system significantly outperforms the baseline and NMF-only systems.

\begin{table} [t]
\caption{\label{table:acc} {\it Overall classification accuracy comparison for the various systems on entire test data}}
\vspace{0.1in}
\centerline{
\begin{tabular}{|c|c|c|}
\hline
 Baseline & NMF-only & Enhanced\\ \hline
 45.1\% & 72.6\% & 78.4\%\\
\hline  
\end{tabular}}
\end{table}

\begin{figure}[t]
\centering
\includegraphics[height=48mm, width = 80mm]{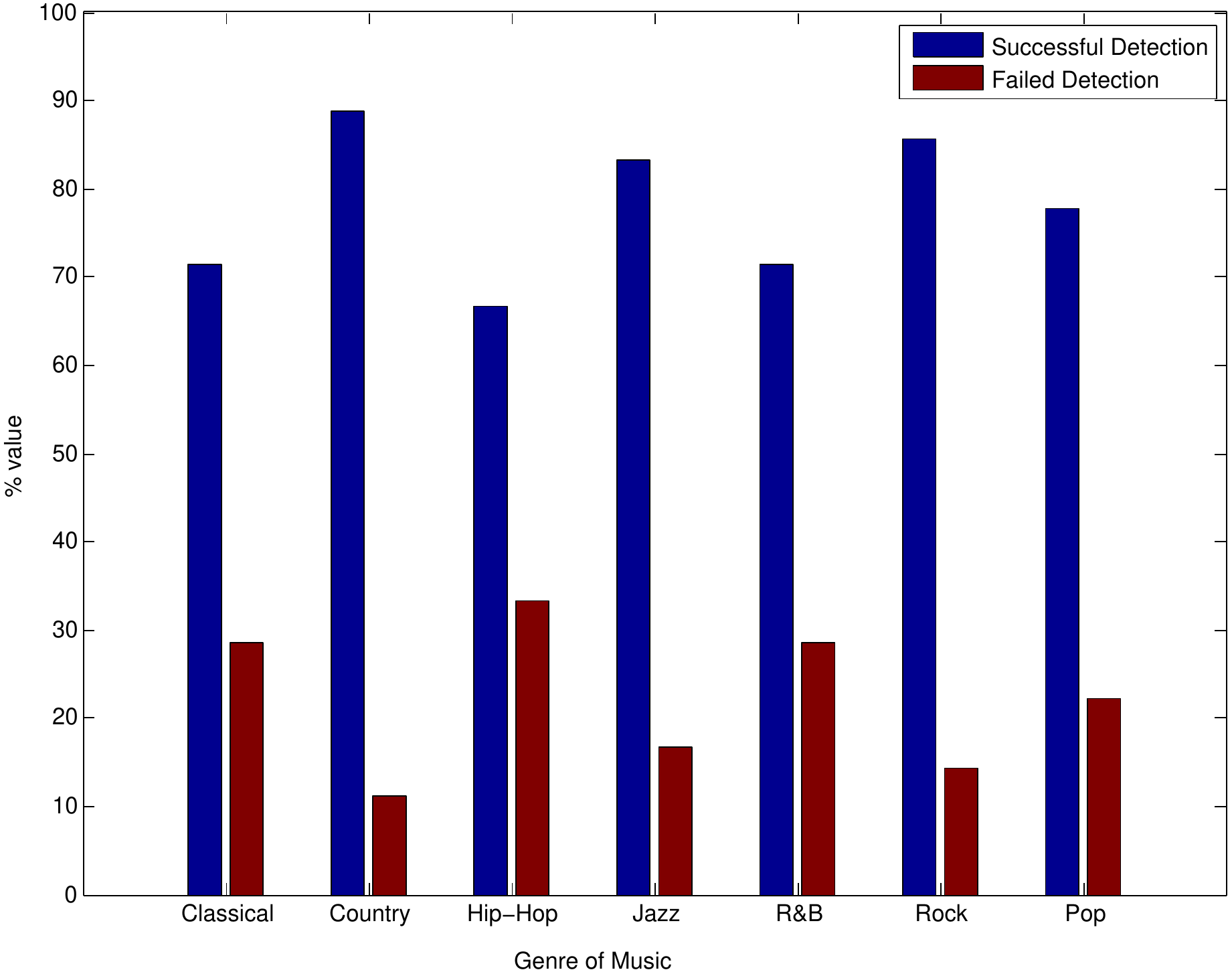}
\caption{\it Performance across genres (Baseline+NMF features)}
\label{fig:genre}
\end{figure}

Figure \ref{fig:genre} shows the successful and failed detection rates for plagiarized track pairs from different genres. Our method performs the best in the country, jazz and rock categories, and worst in hip-hop. This may be because the presence of rap sequences make the detection of these song pairs difficult. 

 An inspection of plagiarized sequences shows that plagiarized pieces often contain only a small sequence that is similar to the original, and our feature set does not do anything to explicitly address this. To account for such cases, one may consider deriving features from the alignment trajectories, to detect local occurrences of systematically varied rhythms.%; {\it e.g.} `Oye Mi Canto' had only a 2 second clip similar to `Paginas De Mujer' in the chorus.

While existing systems for the detection of near-duplicate music documents can be used for plagiarism detection \cite{Hanna2007},  we observe that their performance worsens in polyphonic settings, failing in a number of cases where our method proves successful, {\it e.g.} song pairs {\it He's So Fine} and {\it My Sweet Lord}, {\it Oye Mi Canto} and {\it Paginas De Mujer}. Our method proves least effective in cases where the similar portions are in the background, {\it e.g.} a copied guitar riff in the musical pieces involved in the {\it La Cienega Music Co. v. ZZ Top} plagiarism suit.

\section{Conclusion}
\label{sec:conc}
In this paper, we proposed a novel feature space derived from techniques commonly used in signal separation to account for polyphony in music recordings. We formulated the task of plagiarism detection in a supervised classification framework which using distance features over pairs of music tracks to learn the model. Our approach resulted in a significant improvement in performance over the baseline metrics. It is worth noting that our method does not dispense with the need for agencies that track potential copyright violations. Ideally, it should be used as a filtering mechanism that identifies extremely similar music samples. 

This work used exemplar bases obtained from the data for NMF-- alternate learning methods for the bases (such as training of the basis set from the data, using examples of expected notes for initialization) and effects of changing the basis set size are directions that we expect to explore in future work. 

The NMF-based feature representation should be useful in other tasks that deal with polyphonic music as well. However, for music retrieval tasks such as query by humming or song-matching tasks, we need to be especially careful in the training/selection of basis vectors, because such retrieval tasks are often applied to databases that include user-generated content, typically of a worse quality than studio recordings due to background noise. For such tasks, we would need to have an additional step for noise removal/reduction or ensure that the basis vectors are not trained from studio-quality recordings only. We continue to actively explore these directions.

%\section{Acknowledgements}
%The InterSpeech 2012 organizing committee would like to thank the organizing committee of InterSpeech 2008, 2009, 2010, 2011 for their help and for kindly providing the template files.

%\nocite{*}

%\eightpt
\ninept
\bibliographystyle{unsrt}
\bibliography{Speech_Technology_Paper}
\end{document}